\documentclass[twocolumn]{aastex631}

\newcommand{\eps}{erg s$^{-1}$}
\newcommand{\ecs}{erg cm$^{-2}$ s$^{-1}$}
\newcommand{\pcm}{cm$^{-2}$}
\newcommand{\M}{$M_{\odot}$}
\newcommand{\phc}{ph cm$^{-2}$ s$^{-1}$}

\newcommand{\ecps}{erg cm s$^{-1}$~}
\newcommand{\source}{GRS~1758--258}

\shorttitle{GRS 1758--258}
\shortauthors{Jana et al.}
\graphicspath{{./}{figures/}}
\begin{document}

\title{Broadband X-ray Spectroscopy and Estimation of Spin of the Galactic Black Hole Candidate GRS~1758--258}

\author[0000-0001-7500-5752]{Arghajit Jana}
\affiliation{Institute of Astronomy, National Tsing Hua University, Hsinchu, 30013, Taiwan}

\author[0000-0002-5617-3117]{Hsiang-Kuang Chang}
\affiliation{Institute of Astronomy, National Tsing Hua University, Hsinchu, 30013, Taiwan}

\author[0000-0003-3932-6705]{Arka Chatterjee}
\affiliation{Department of Physics and Astronomy, University of Manitoba, Manitoba, R3T 2N2, Canada}

\author[0000-0003-2865-4666]{Sachindra Naik}
\affiliation{Astronomy \& Astrophysics Division, Physical Research Laboratory, Navrangpura, Ahmedabad, 380009, India}

\author[0000-0001-6189-7665]{Samar Safi-Harb}
\affiliation{Department of Physics and Astronomy, University of Manitoba, Manitoba, R3T 2N2, Canada}



\begin{abstract}
We present the results of a broadband ($0.5-78$~keV) X-ray spectral study of the persistent Galactic black hole X-ray binary GRS~1758--258 observed simultaneously by Swift and NuSTAR. Fitting with an absorbed power-law model revealed a broad Fe line and reflection hump in the spectrum. We used different flavours of the relativistic reflection model for the spectral analysis. All models indicate the spin of the black hole in GRS~1758--258 is $>0.92$. The source was in the low hard state during the observation, with the hot electron temperature of the corona  estimated to be $kT_{e} \sim 140$~keV. The black hole is found to be accreting at $\sim 1.5\%$ of the Eddington limit during the observation, assuming the black hole mass of 10 $M_{\odot}$ and distance of $8$~kpc.
\end{abstract}



\section{Introduction} \label{sec:intro}
Black hole X-ray binaries (BHXRBs) are powered by the accretion process, where matter from the companion star get accreted on to the central black hole (BH). The gravitational energy is converted to radiation emitted over the electromagnetic spectrum in the accretion process. Depending on the X-ray activity, a BHXRB can be classified into a transient or a persistent source \citep{Tetarenko2016}. The transient source spends most of the time in a quiescent state with very low X-ray luminosity ($L_{\rm X} <10^{32}$ \eps) and occasionally undergoes an outbursting phase when the X-ray luminosity increases to $L_{\rm X} > 10^{35}$ \eps. On the other hand, a persistent source is always found to be active in X-rays with the X-ray luminosity $L_{\rm X} \sim 10^{35-37}$ \eps \citep[e.g.,][]{Tetarenko2016}.

A black hole is characterized by its mass ($M_{\rm BH}$) and spin ($a^*$). Estimation of the BH spin is harder compared to the estimation of BH mass. There exist various direct methods to estimate the BH mass, such as radial velocity measurement of the secondary, dips and eclipses in the light curves \citep[e.g.,][]{Kreidberg2012,Torres2019,Jana2022}. The BH mass can also be estimated from the spectral modelling and timing analysis \citep[e.g.,][]{Kubota1998,Shaposhnikov2007,Shaposhnikov2009,AJ2016,AJ2020b,AJ2021c,DC2016}. The BH spin can be measured from X-ray spectroscopy. In this method, the spin can be estimated in two processes: the continuum fitting (CF) method \citep[CF; e.g.,][]{Zhang1997,McClintock2006,Steiner2014}, and the study of Fe line and reflection spectroscopy \citep[e.g.,][]{Fabian1989,Miller2012,Reynolds2020}. Both methods require measuring the inner edge of the accretion disk that extends up to the inner most stable circular orbit (ISCO). It has also been suggested that the black hole spin can affect the polarization state of X-ray emission \citep[e.g.,][]{Schnittman2010,Dovciak2011}.

In the CF method, the inner radius of the accretion disk is measured by fitting the thermal disk continuum with a general relativistic model \citep[e.g.,][]{Gierlinski2001,McClintock2006,Shafee2006}. In this process, one also needs to have prior knowledge of the BH mass, distance and inclination angle. On the contrary, no knowledge of BH mass or distance is required to measure the spin in reflection spectroscopy. In this method, the reflection of the coronal emission at the inner disk is studied. The essential features of the reflection spectra are Fe fluorescent emission between $6.4-6.97$~keV and a reflection hump around $15-40$~keV. The accretion disk is optically thick and geometrically thin and extends up to the ISCO \citep{SS73}. Due to the relativistic effect (Doppler shifts and gravitational redshift), the line profile of the Fe line originating from the inner accretion disk is blurred. As the ISCO depends on the spin, the study of the blurred spectra allows us to estimate the spin of the BH \citep{Bardeen1972,Novikov1973}. 

The spin of black hole has been estimated using either CF method or reflection spectroscopy, or both. The CF method has been used to estimate spin for LMC~X--1 \citep{Mudambi2020}, LMC~X--3 \citep{Bhuvana2021}, MAXI~J1820+070 \citep{Zhao2021}. The reflection spectroscopy is used to estimate spin for several BHs, e.g., MAXI~J1535--571 \citep{Miller2018}, XTE~J1908--094 \citep{Draghis2021}, Cygnus X--1 \citep{Tomsick2018}, MAXI~J1631--479 \citep{Xu2020}, GX~339--4 \citep{Garcia2019}. Both reflection spectroscopy and CF have been used to measure the spin in a few BHs, e.g., GRS~1716--249 \citep{Tao2019}, LMC~X--3 \citep{AJ2021d}, GX~339--4 \citep{Parker2016}. The majority of the BHs are found to have prograde spin, i.e., the accretion flow rotates in the same direction as the BH . Only a few BHs are found to have a retrograde spin, e.g., MAXI~J1659--152 \citep{Rout2020}, Swift~J1910.2-0546 \citep{Reis2013}, GS~1124--683 \citep{Morningstar2014}. Nonetheless, the spin of the BH has been observed to have a wide range. Although, the spin has been measured for a substantial number of BHs, the spin of many BH remains unknown.

GRS~1758--258 is a black hole X-ray binary located in the close vicinity of the Galactic centre. GRS~1758--258 was discovered by {\it GRANAT}/SIGMA in 1990 \citep{Syunyaev1991}. It is one of the few persistent BHXRBs in our galaxy. The source has been observed in multi-wavelengths over the years \citep[e.g.,][]{Rodriguez1992,Mereghetti1994,Mereghetti1997,Smith2001,Keck2001,Pottschmidt2006,Lin2000,Luque2014}. It is considered to be a black hole based on its spectral and timing properties \citep{Sidoli2002}. GRS~1758--258 is predominately found to be in the low hard state \citep[LHS;][]{Soria2011}. From the RXTE monitoring, \citet{Smith2002} reported a state transition to the soft state in 2001 with the $3-25$~keV flux decreased by over an order of magnitude. GRS~1758--258 also shows two extended radio lobes, which makes the system a microquasar \citep{Rodriguez1992}. For a long time, the companion star was not identified due to the dense stellar population in the field. Recently, the spectroscopic study suggested that the companion is likely an A-type main-sequence star \citep{Marti2016}. The orbital period of GRS~1758--258 is reported to be $18.45\pm0.10$~days \citep{Smith2002}.

GRS~1758--258 is the least studied source among three persistent Galactic black hole binaries. The mass and spin of the black hole in GRS~1758--258 are not known yet. In this paper, we aim to estimate the spin of the BH in GRS~1758--258 from the broadband X-ray spectroscopy. The paper is organized in the following way. We present the observation and data reduction technique in \S2. In \S3, we present the analysis and result. Finally, we discuss our findings in \S4.

\begin{table*}
\centering
\caption{Log of Observations of GRS~1758--258}
\begin{tabular}{lccccc}
\hline
Instrument & Date (UT) & Obs ID & Exposure (ks)  & Count s$^{-1}$\\
\hline
NuSTAR & 2018-09-28 & 30401030002 & 42  & $17.39\pm0.02$\\
Swift/XRT   & 2018-09-28 & 00088767001 & 1.7 & $ 7.88\pm0.07$\\
\hline
\end{tabular}
\label{tab:log}
\end{table*}

\section{Observations and Data Reduction}
NuSTAR observed \source~ on September 28, 2018 for a total exposure of 42~ks (see Table~\ref{tab:log}). NuSTAR is a hard X-ray focusing telescope, consisting of two identical modules: FPMA and FPMB \citep{Harrison2013}. The raw data were reprocessed with the NuSTAR Data Analysis Software ({\tt NuSTARDAS}, version 1.4.1). Cleaned event files were generated and calibrated by using the standard filtering criteria in the {\tt nupipeline} task and the latest calibration data files available in the NuSTAR calibration database (CALDB) \footnote{\url{http://heasarc.gsfc.nasa.gov/FTP/caldb/data/nustar/fpm/}}. The source and background products were extracted by considering circular regions with radii 60 arcsec and 90 arcsec, at the source co-ordinate and away from the source, respectively. The spectra and light curves were extracted using the {\tt nuproduct} task. We re-binned the spectra with 30 counts per bin by using the {\tt grppha} task. 

Swift/XRT observed \source~ simultaneously with NuSTAR for an exposure of 1.7 ks in window-timing (WT) mode. The XRT spectrum did not suffer from photon pile-up. In general, the pile-up occurs if the count rate is over 100 counts s$^{-1}$ in the WT mode \citep{Romano2006}. The $0.5-10$~keV spectrum was generated using the standard online tools\footnote{\url{http://www.swift.ac.uk/user_objects/}} provided by the UK {\it Swift} Science Data Centre \citep{Evans2009}. For the present study, we used simultaneous observations of Swift/XRT and NuSTAR in the $0.5-78$~keV energy range.

\begin{figure}
\centering
\includegraphics[width=8.5cm]{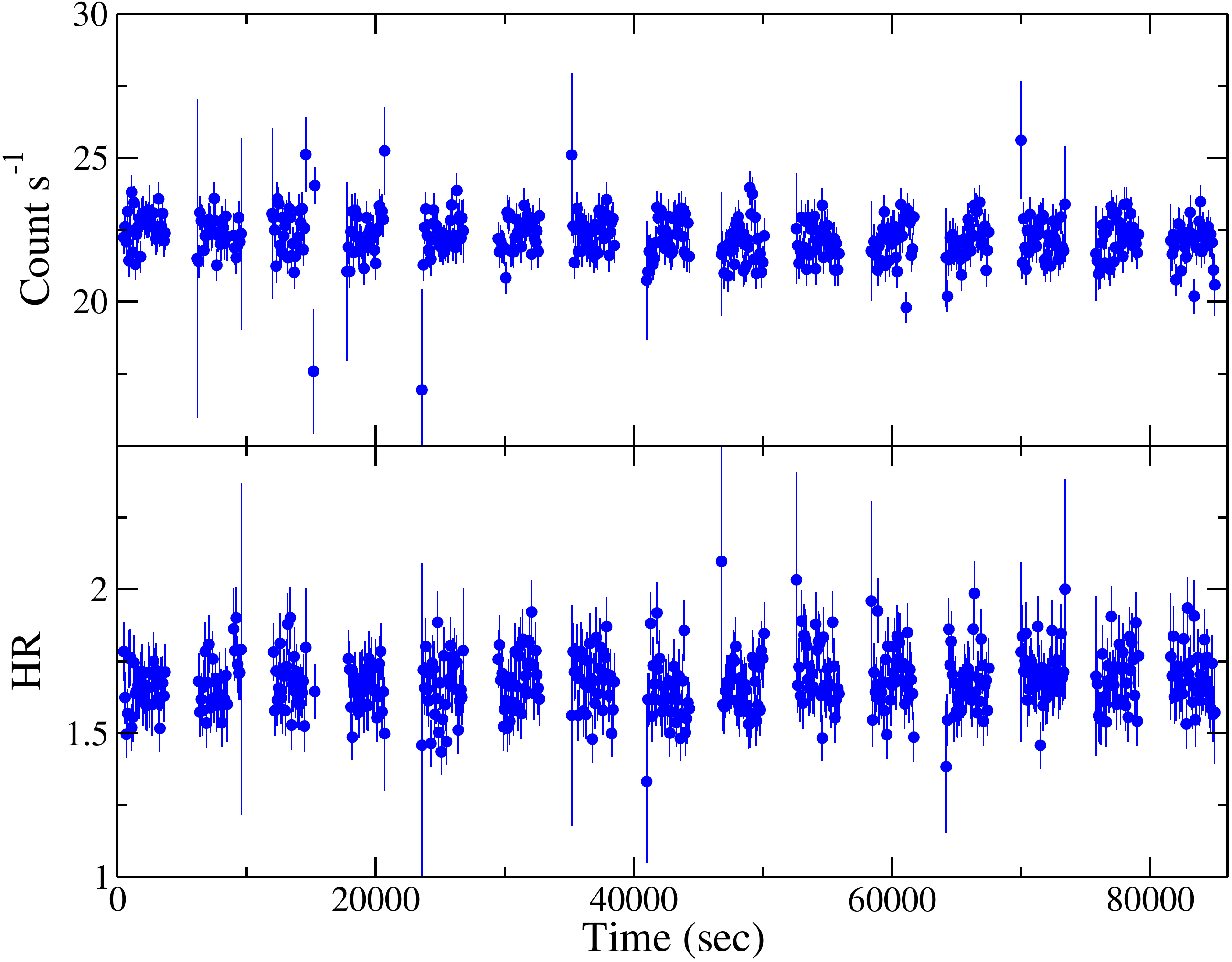}
\caption{Top panel: $3-78$~keV light curve of \source~ from the NuSTAR observation. The light curve is binned at 100s. Bottom panel: Variation of the hardness ratio (HR). The HR is defined as the ratio of count rates in the $6-30$~keV range to the $3-6$~keV range.}
\label{fig:lc}
\end{figure}

\section{Analysis and Result}
\label{sec:analysis}

Figure~\ref{fig:lc} shows the $3-78$~keV light curve of \source~ from the NuSTAR observation in the top panel. In the bottom panel of Figure~\ref{fig:lc}, the variation of the hardness ratio (HR) is shown. We define the HR as the ratio of the count rate in the $6-30$~keV to the $3-6$~keV energy ranges. We did not observe any variation in the count rate or HR during the observation period. Hence, we carried out the spectral analysis using the time averaged spectrum obtained from the observation. The spectral analysis was carried out in {\tt HEASEARC}'s spectral analysis package {\tt XSPEC} version 12.8.2 \citep{Arnaud1996}. For the analysis, we used \textsc{tbabs} model for the interstellar absorption with the \textsc{wilms} abundance \citep{Wilms2000} and the cross-section that of \citet{Verner1996}.

\begin{figure}
\centering
\includegraphics[width=8.5cm]{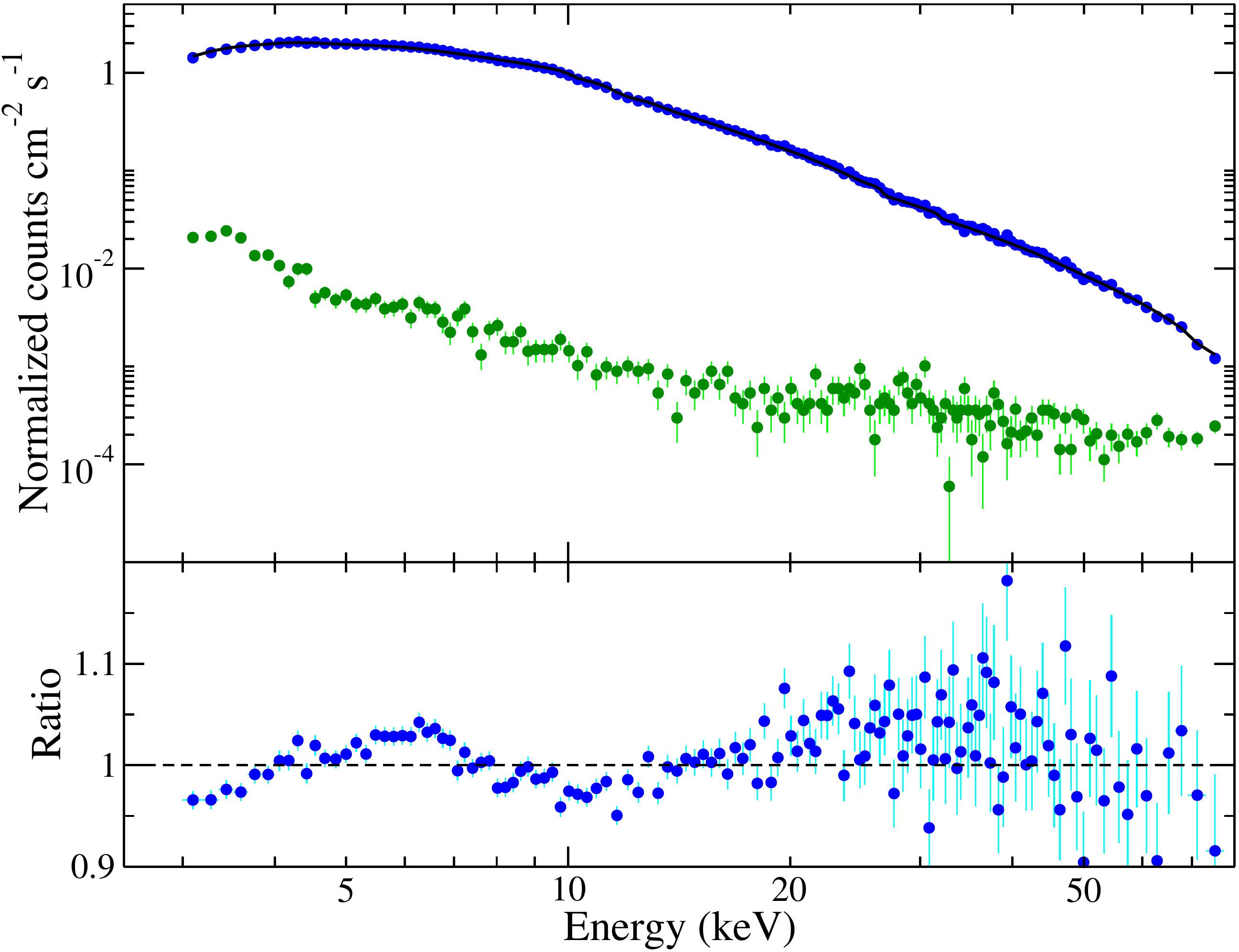}
\caption{Top panel : \textsc{cutoff} model fitted NuSTAR spectrum in the $3-78$~keV energy range. The blue and green points represent the source and background count, respectively. Bottom panel : the residual in terms of data/model ratio.}
\label{fig:nu-spec}
\end{figure}

\begin{figure}
\centering
\includegraphics[width=8.5cm]{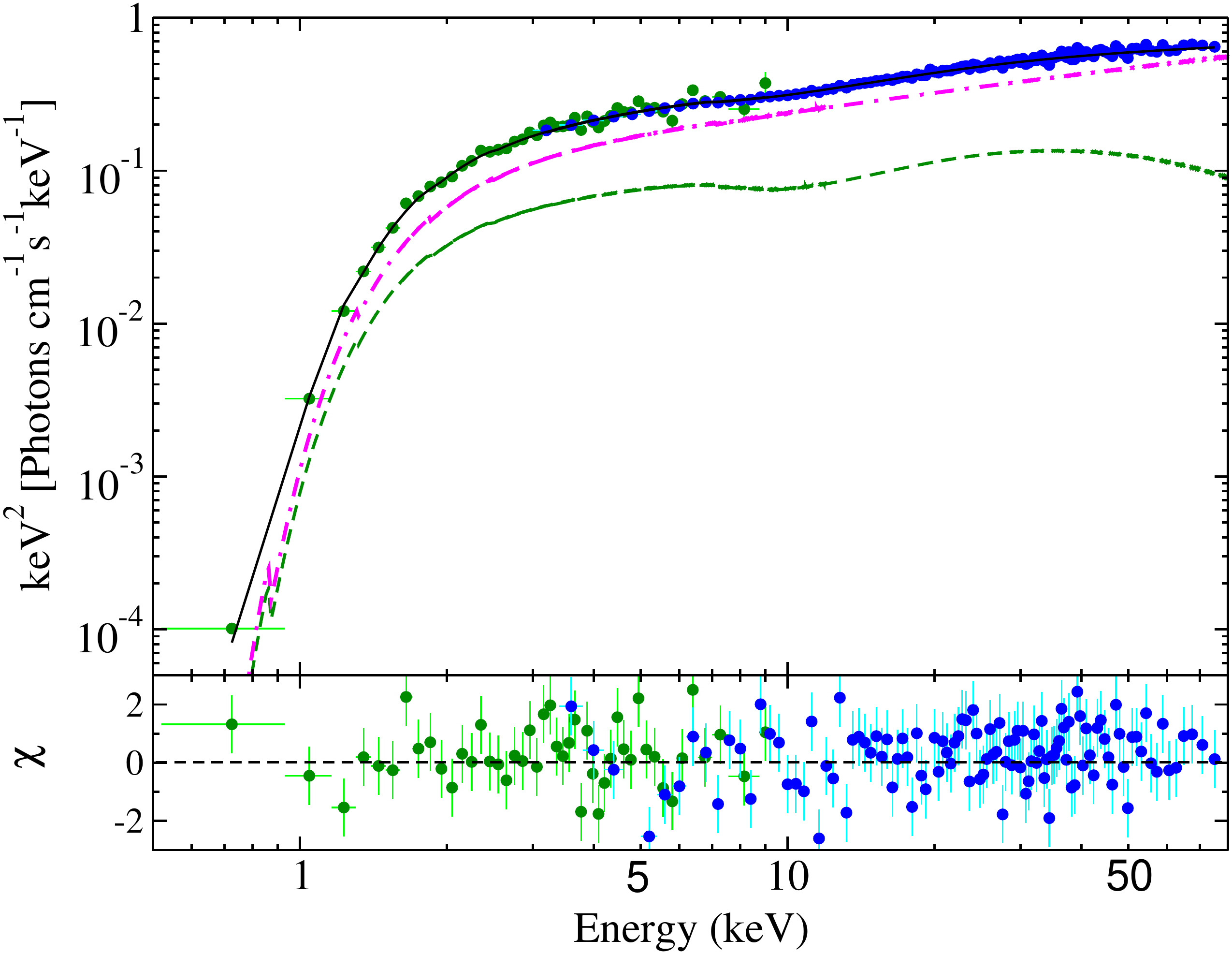}
\caption{Top: \textsc{RelxillLp} model fitted $0.5-78$~keV XRT+NuSTAR unfolded spectrum. The dot-dashed magenta and dashed green lines represent the primary and reprocessed emission components, respectively.} Bottom: Corresponding residual in terms of data-model/error. The green and blue points represent the XRT and NuSTAR data, respectively.
\label{fig:spec}
\end{figure}

\begin{figure}
\centering
\includegraphics[width=8.5cm]{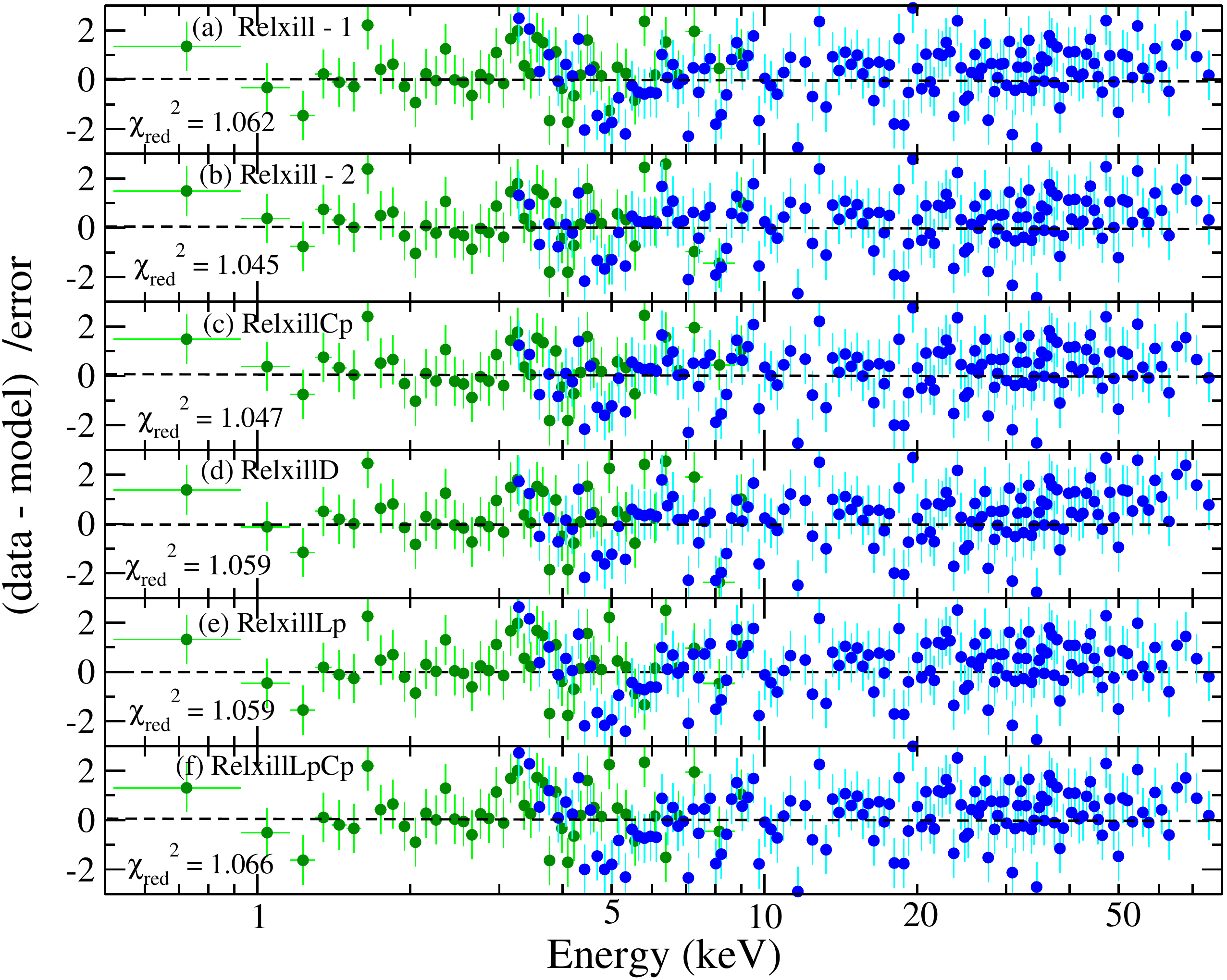}
\caption{Residuals in terms of (data-model)/error obtained from the spectral analysis with (a) \textsc{Relxill-1}, (b) \textsc{Relxill-2}, (c) \textsc{RelxillCp}, (d) \textsc{RelxillD}, (e) \textsc{RelxillLp} and (f) \textsc{RelxillLpCp} models. The green and blue points represent the XRT and NuSTAR data, respectively. Corresponding reduced-$\chi^2$ is quoted in the inset of each panel.}
\label{fig:del}
\end{figure}

Generally, X-ray spectra of a BHXRBs can be approximated by a multi-colour disk blackbody (MCD) and power-law components.  Additionally, reprocessed emission may be observed : a Fe K$\alpha$ line at $\sim 6.4$~keV and a reflection hump at $\sim 15-40$~keV. We started our spectral analysis with the Swift/XRT+NuSTAR data in the $0.5-78$~keV energy range. The MCD component may not be present in the LHS. Thus, we attempted to fit the data with the absorbed power-law model with an exponential high energy cutoff. This model did not give us an acceptable fit with clear signatures of Fe K line, and a reflection hump in the $5-8$~keV and $15-40$~keV energy ranges, respectively. Figure~\ref{fig:nu-spec} shows the $3-78$~keV NuSTAR spectrum in the top panel. For clarity, we only show the NuSTAR spectrum. The bottom panel of Figure~\ref{fig:nu-spec} shows the residuals in terms of data/model ratio. We added a Gaussian line to incorporate Fe K-line in the $5-8$~keV energy range to improve the fit. Although, the fit improved with $\Delta \chi^2 =91$ for 3 degrees of freedom (dof), it was still unacceptable as the reflection hump was clearly visible in the residuals at energies above $\sim 10$ keV.

To probe the reprocessed emission, we employed the relativistic reflection model \textsc{Relxill} \citep{Garcia2013,Garcia2014,Dauser2014,Dauser2016} for further spectral analysis. We applied different flavours of the \textsc{Relxill} model with different assumptions to the $0.5-78$~keV XRT+NuSTAR spectra. We used different \textsc{Relxill} family of models, namely, \textsc{Relxill}, \textsc{RelxillCp}, \textsc{RelxillD}, \textsc{RelxillLp} and \textsc{RelxillLpCp} for the spectral analysis. Figure~\ref{fig:spec} shows the \textsc{RelxillLp} model fitted XRT+NuSTAR spectrum in the $0.5-78$~keV energy range in the top panel. The dot-dashed magenta and dashed green lines represent the primary emission and reprocessed emission, respectively. Corresponding residuals in terms of data/model ratio are shown in the bottom panel. Figure~\ref{fig:del} shows the residuals obtained from the different models in the different panels. The green and blue points represent the XRT and NuSTAR data, respectively. In the inset of each panel, the value of corresponding reduced-$\chi^2$ is mentioned.

\subsection{Relxill}
\textsc{Relxill} model uses a cutoff power-law as an incident primary emission. In this model, the reflection strength is measured in terms of reflection fraction ($R_{\rm refl}$) which is defined as the ratio of reflected emission to the direct emission to the observer. A broken power-law emission profile is assumed in \textsc{Relxill} model with $E(r) \sim R^{-q_{\rm in}}$ for $r > R_{\rm br}$ and $E(r) \sim R^{-q_{\rm out}}$ for $r < R_{\rm br}$, where E(r), $q_{\rm in}$, $q_{\rm out}$ and $R_{\rm br}$ are emissivity, inner emissivity index, outer emissivity index and break radius, respectively.

We used \textsc{Relxill} model in two assumptions: first by keeping the inner and outer emissivity indices fixed at the default value, i.e. $q_{\rm in} = q_{\rm out} = 3$ (hereafter Relxill-1); and second, allowing the inner and outer emissivity index to vary freely (hereafter Relxill-2). During our analysis, we fixed the outer disk at $R_{\rm out} = 1000~R_{\rm g}$. Both models gave us a good fit with $\chi^2$/dof = 1718/1618 and 1691/1617 for Relxill-1 and Relxill-2, respectively, although \textsc{Relxill-2} gave us a better fit. Both models indicated a high spinning black hole with spin parameter, $a^* > 0.94$. The accretion disk is found to extend almost up to the ISCO, as we obtained $R_{\rm in} = 1.14^{+0.04}_{-0.03}$ $R_{\rm ISCO}$ and  $1.13^{+0.02}_{-0.04}$ $R_{\rm ISCO}$ from Relxill-1 and Relxill-2, respectively. We could not constrain the cutoff energy ($E_{\rm cut}$) form these two models, as the cutoff energy pegged at the upper limit of the model at $1000$~keV. The reflection is found to be weak with $R_{\rm refl} \sim 0.23$ and $\sim 0.21$ from \textsc{Relxill-1} and \textsc{Relxill-2} models, respectively. With \textsc{Relxill-2}, we obtained a steep inner emissivity ($q_{\rm in}=6.78^{+0.05}_{-0.07}$) and a flat outer emissivity index ($q_{\rm out}=2.04^{+0.09}_{-0.06}$) with the break radius $R_{\rm br}=4.8^{+1.5}_{-0.8}$ $R_g$.

We also included a disk component in our spectral analysis to check if the disk is present. The thermal disk component was modelled with \textsc{diskbb} \citep{Makishima1986}. We obtained the inner disk temperature, $kT_{\rm in}=0.18\pm0.12 $~keV. The other spectral parameters remained the same. The addition of the disk component did not improve our fit for both \textsc{Relxill-1} and \textsc{Relxill-2} models. We checked if the disk component is required with {\tt FTOOLS} task \textsc{ftest}. The \textsc{ftest} returned with probability=1, indicating the disk component was not required. As statistically, the disk component was not required; hence, we did not add the disk component for further analysis.

\subsection{RelxillCp}
For further spectral analysis, we applied the \textsc{RelxillCp} model for the spectral analysis. The \textsc{RelxillCp} has advantage over \textsc{Relxill}, as the \textsc{RelxillCp} directly estimate the coronal properties, namely, the hot electron plasma temperature ($kT_{\rm e}$). In \textsc{RelxillCp}, the primary incident spectrum is computed using \textsc{nthcomp} \citep{Z96,Zycki1999} model, replacing \textsc{cutoffpl} in the \textsc{Relxill} model. The analysis with the \textsc{RelxillCp} returned with a good fit with $\chi^2=1691$ for 1617 dof. We obtained a similar fit with this model with the spin and inner disk radius were found to be, $a^* = 0.97^{+0.01}_{-0.02}$ and $R_{\rm in} = 1.13^{+0.03}_{-0.03}$ $R_{\rm g}$, respectively. We also obtained the temperature of the Compton corona as $kT_e = 134^{+82}_{-29}$~keV. 

\subsection{High Density Model: RelxillD}
\textsc{Relxill} and \textsc{RelxillCp} models consider a fixed density of the accretion disk as $n=10^{15}$ cm$^{-3}$, while \textsc{RelxillD} allows the density to vary. In this model, the incident primary emission is a cutoff power-law with cutoff energy fixed at 300~keV. The spectral analysis with the \textsc{RelxillD} model returned with a high spin $a^* > 0.98$. We obtained a steeper emissivity with \textsc{RelxillD} model with $q_{\rm in}=7.09^{+0.07}_{-0.11}$, compared to the \textsc{Relxill} and \textsc{RelxillCp} models. The inclination angle of the inner disk is found to be higher with higher density, with $i=37^{+2}_{-3}$ degrees. We obtained the disk density as $n<2\times10^{15}$ cm$^{-3}$. The detailed spectral analysis result is tabulated in Table~\ref{tab:my_label}.

\subsection{Lamppost Geometry: RelxillLp and RelxillLpCp}
In the \textsc{Relxill} model, no particular geometry of the corona is assumed. In the lamp post model, the corona is assumed to be a point source, located above the BH \citep{Garcia2010,Dauser2016}. \textsc{RelxillLp} and \textsc{RelxillLpCp} flavors of \textsc{Relxill} family of models assumed the lamp post geometry. The incident primary emission is either \textsc{cutoff} (\textsc{RelxillLp}) or \textsc{nthcomp} (\textsc{RelxillLpCp}). The height of corona ($h$) is an input parameter in this model.

We obtained a good fit with both \textsc{RelxillLp} and \textsc{RelxillLpCp} models, with the fit returned as $\chi^2=1715$ and $\chi^2=1726$ for 1619 dof, respectively. The coronal height ($h$) is obtained to be $h=3.4^{+1.1}_{-0.7}$ $R_g$ and $h=3.7^{+0.9}_{-0.6}$ $R_g$ from \textsc{RelxillLp} and \textsc{RelxillLpCp} models, respectively. The spin of the BH is obtained to be $a^*<0.92$ from the analysis with these models. The reflection is obtained to be stronger with $R_{\rm refl}\sim 0.46$, compared to the other models. Table~\ref{tab:my_label} shows the detailed spectral analysis results with the \textsc{RelxillLp} and \textsc{RelxillLpCp} models.

\subsection{Error Estimation}
\begin{figure*}
\centering
\includegraphics[height=12cm,width=18cm]{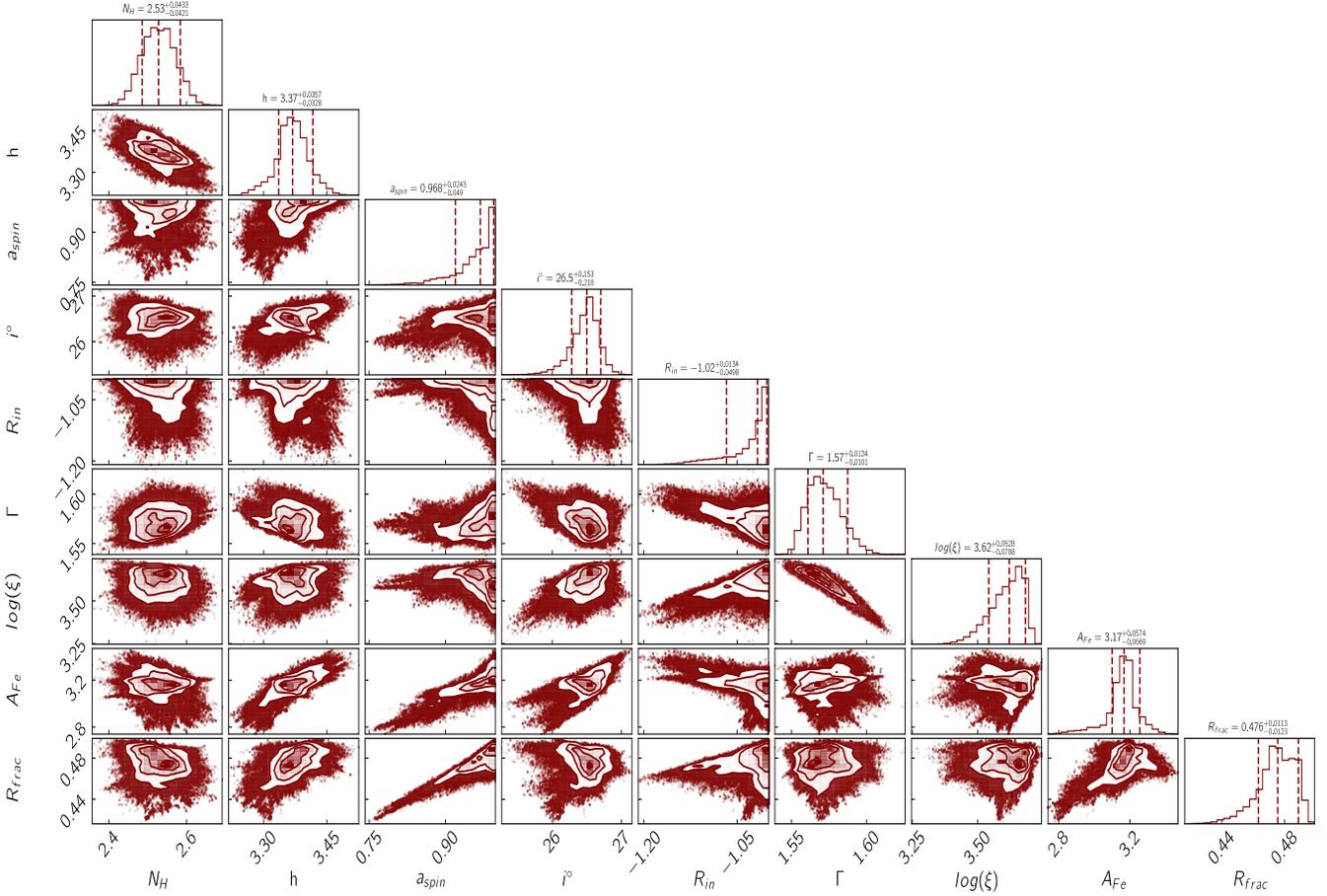}
\caption{Posterior distribution of the spectral parameters obtained from the MCMC analysis with the \textsc{RelxillLp} model. Plotting was performed using \texttt{corner} \citep{corner} plot. Central dashed lines correspond to the peak values whereas $1\sigma$ confidence levels are represented by dashed lines on either sides.}
\label{fig:mcmc}
\end{figure*}

To calculate the uncertainty of the best-fitted spectral parameters, we ran Monte Carlo Markov Chain (MCMC) in {\tt XSPEC}\footnote{\url{https://heasarc.gsfc.nasa.gov/xanadu/xspec/manual/node43.html}}. The chains were run with eight walkers for a total of $10^6$ steps using the Goodman-Weare algorithm. We discarded the first 10000 steps of the chains, assuming to be in the `burn-in' phase. The uncertainty is estimated with `error' command at  1.6$\sigma$ confidence level. In the paper, we used the error at 1.6$\sigma$ level (90\% confidence), or mentioned otherwise. Figure~\ref{fig:mcmc} shows the posterior distribution of the spectral parameters and errors obtained with the \textsc{RelxillLp} model. The errors are mentioned at 1 $\sigma$ level in Figure~\ref{fig:mcmc}. Following models, like \textsc{relxill}, \textsc{relxillCp} or \textsc{relxillD} do not assume any coronal geometry, while \textsc{relxillLp} assumes lamp-post geometry. As the \textsc{relxillLp} provides a physical picture of the coronal geometry compared to other models mentioned before, we used \textsc{RelxillLp} model for the MCMC analysis.

\begin{table*}
\centering
\caption{Spectral Analysis results}
\label{tab:my_label}
\begin{tabular}{lccccccccc}
\hline
 & CUTOFF &  Relxill-1 & Relxill-2 & RelxillCp & RelxillD & RelxillLp & RelxillLpCp \\
\hline
$N_{\rm H}$ ($10^{22}$ \pcm)& $0.86^{+0.18}_{-0.18}$   & $2.55^{+0.01}_{-0.01}$ &  $2.56^{+0.02}_{-0.01}$ & $2.54^{+0.02}_{-0.02}$ & $2.53^{+0.02}_{-0.01}$ & $2.53^{+0.03}_{-0.05}$ & $2.52^{+0.02}_{-0.02}$\\
\\
$\Gamma$    & $1.56^{+0.02}_{-0.01}$   & $1.54^{+0.01}_{-0.01}$ & $1.53^{+0.02}_{-0.02}$ & $1.56^{+0.02}_{-0.02}$ & $1.54^{+0.02}_{-0.01}$ & $1.57^{+0.01}_{-0.02}$ & $1.53^{+0.02}_{-0.01}$\\
\\
$E_{\rm cut}$/$kT_{e}$ (keV) & $500^{+u}_{-28}$  & $>946$ & $>925$ & $134^{+82}_{-29}$  & $300^{\dagger}$ & $>911$ & $146^{+68}_{-32}$ \\
\\
$q_{\rm in}$ & -- & $3^{f}$ & $6.78^{+0.05}_{-0.07}$ & $6.51^{+0.06}_{-0.05}$ & $7.09^{+0.07}_{-0.11}$ & $-$ & $-$ \\
\\
$q_{\rm out}$ & -- & $3^f$ & $2.04^{+0.09}_{-0.06}$  & $1.99^{+0.08}_{-0.07}$ & $2.10^{+0.04}_{-0.06}$ & -- & --\\
\\
$R_{\rm br}$/$h$ ($R_g$) & --   & $10.8^{+2.5}_{-1.4}$ & $4.8^{+1.5}_{-0.8}$ &  $5.3^{+1.7}_{-2.2}$ & $4.5^{+1.6}_{-0.8}$ & $3.4^{+0.6}_{-0.4}$ & $3.7^{+0.9}_{-0.6}$ \\
\\
$R_{\rm in}$ ($R_{\rm ISCO}$) & --  & $1.14^{+0.04}_{-0.03}$ & $1.13^{+0.02}_{-0.04}$ & $1.13^{+0.03}_{-0.03}$ & $1.15^{+0.04}_{-0.03}$ &  $<1.04$ & $<1.04$\\
\\
$a^*$ & -- &   $0.97^{+0.02}_{-0.01}$ & $0.96^{+0.02}_{-0.02}$ & $0.97^{+0.01}_{-0.02}$ & $>0.98$ & $0.97^{+0.02}_{-0.05}$ & $0.95^{+0.01}_{-0.03}$\\
\\
$i$ (degree) & --  & $31^{+2}_{-2}$ & $32^{+2}_{-4}$ & $29^{+2}_{-3}$ & $37^{+2}_{-3}$ & $27^{+2}_{-1}$ & $28^{+2}_{-2}$\\
\\
$\log(n)$ (log cm$^{-3}$) & -- & $15^{\dagger}$ &  $15^{\dagger}$ &  $15^{\dagger}$ &  $<15.3$ &  $15^{\dagger}$ &  $15^{\dagger}$ & \\
\\
$\log \xi$ (log \ecps)& -- &  $3.68^{+0.03}_{-0.07}$ & $3.61^{+0.10}_{-0.12}$ & $3.71^{+0.14}_{-0.11}$ & $3.90^{+0.07}_{-0.15}$ & $3.62^{+0.10}_{-0.12}$ & $3.72^{+0.10}_{-0.13}$ \\
\\
$A_{\rm Fe}$ ($A_{\sun}$)& -- & $2.75^{+0.07}_{-0.13}$ & $3.16^{+0.19}_{-0.13}$ & $3.16^{+0.10}_{-0.14}$ & $2.99^{+0.11}_{-0.15}$ & $3.17^{+0.11}_{-0.13}$ & $3.28^{+0.12}_{-0.11}$ \\
\\
$R_{\rm refl}$ & --  & $0.31^{+0.02}_{-0.04}$ & $0.23^{+0.02}_{-0.01}$  & $0.21^{+0.04}_{-0.03}$ & $0.16^{+0.02}_{-0.01}$ & $0.48^{+0.03}_{-0.04}$ & $0.46^{+0.03}_{-0.02}$\\
\\
$N_{\rm PL}$/$N_{\rm rel}$ ($10^{-3}$ \phc) & $0.12^{+0.01}_{-0.01}$  & $5.02^{+0.05}_{-0.03}$  & $4.75^{+0.06}_{-0.05}$ & $4.68^{+0.05}_{-0.07}$ & $4.61^{+0.03}_{-0.04}$ & $22.7^{+0.7}_{-0.5}$ & $22.7^{+0.5}_{-0.8}$\\
\\
$\chi^2$/dof & 1484/1165  & 1718/1618 & 1691/1617 & 1691/1617 & 1712/1617 & 1715/1619 & 1726/1619\\
\\
$\chi^2_{\rm red}$ & 1.273  & 1.062 & 1.045 & 1.047 & 1.059 & 1.059 & 1.066 \\
\\
$F_{\rm 2-10~keV}$ & $5.49^{+0.04}_{-0.03}$  & $5.69^{+0.04}_{-0.05}$ & $5.71^{+0.03}_{-0.04}$ & $5.70^{+0.02}_{-0.03}$ & $5.64^{+0.06}_{-0.05}$ & $5.68^{+0.03}_{-0.04}$  & $5.68^{+0.04}_{-0.05}$ \\
\\
\hline
\end{tabular}
\leftline{$^*$ indicate fixed value in the model. $^f$ indicate the value is fixed during the analysis.}
\leftline{The $F_{\rm 2-10~keV}$ is in unit of $10^{-10}$~ergs cm$^{-2}$ s$^{-1}$.}
\leftline{Errors are quoted at 1.6$\sigma$.}
\end{table*}

\section{Discussion and Concluding Remarks}
We studied the spectral properties of \source~ using the data obtained by NuSTAR and Swift observatories in the energy range of $0.5-78$~keV. We used various spectral models to understand the inner accretion flow during the observation period.

Figure~\ref{fig:mcmc} shows the degeneracy between several parameters. The spin parameter ($a*$) is observed to be correlated with the reflection fraction ($R_{\rm refl}$) and anti-correlated with the inner disk radius ($R_{\rm in}$). These are expected as the high spin would bring the disk closer to the BH, and would make reprocessed emission strong. We also observed the photon index ($\Gamma$) is anti-correlated with the ionization parameter ($\xi$). This indicates if the photon index decreases, the ionization increases which means the hard X-rays are more effective in ionizing the disk.

During our analysis, the hydrogen column density was obtained to be $N_{\rm H} \sim 2.5 \times 10^{22}$ \pcm. However, this is higher than the previously reported $N_{\rm H}$. \citet{Soria2011} found that $N_{\rm H} \sim 1.5 \times 10^{22}$ \pcm. The discrepancy arises due to consideration of the different abundances during the spectral analysis. \citet{Soria2011} assumed the abundances to be \textsc{angr} \citep{Anders1989}, while we assumed \textsc{wilm} abundances. We checked this by assuming \textsc{angr} abundances during the spectral fits. Using \textsc{angr} abundances, the column density was obtained to be $N_{\rm H} \sim 1.5 \times 10^{22}$ \pcm.

The unabsorbed flux was obtained to be $F_{\rm 2-10~keV} \sim 5.7 \times 10^{-10}$ \ecs~ in the $2-10$~keV energy range. The bolometric flux ($F_{\rm bol}$; in the 0.1--500 keV energy band) is estimated to be $F_{\rm bol} \sim 2.6 \times 10^{-9}$ \ecs. From this, we calculated the bolometric luminosity as, $L_{\rm bol} \sim 2\times 10^{37}$ \eps, assuming the source distance of 8~kpc \citep{Soria2011}. Considering the mass of the BH in \source~ as 10 $M_{\odot}$ \citep{Soria2011}, the Eddington ratio is obtained as $L/L_{\rm Edd} \sim 0.015$ or $1.5\%$ of the Eddington limit during the NuSTAR observation. GRS~1758--258 is predominately observed in a similar accretion state with the X-ray luminosity $L_{\rm X} \sim 0.01-0.03$ $L_{\rm Edd}$ \citep{Soria2011}.

The spectral analysis indicated that the source was observed in the LHS. The spectrum was found to be hard, with the photon index $\Gamma \sim 1.53-1.57$ from different spectral models. We could not estimate the cutoff energy as it was not constraint. The spectral fits with the \textsc{RelxillCp} gave us the temperature of the Corona as $kT_{\rm e} = 134^{+82}_{-29}$~keV. Assuming lamp-post geometry, we obtained the corona temperature as $kT_{\rm e}=146^{+68}_{-32}$~keV. For completeness, we calculate the optical depth ($\tau$) of the corona using, $\tau=\sqrt{\frac{9}{4}+\frac{m_{\rm e}c^2}{kT_{\rm e}} \frac{3}{(\Gamma-1)(\Gamma+2)}}-\frac{3}{2}$ \citep{Z96}. The optical depth is obtained to be $\tau = 1.33^{+0.31}_{-0.54}$ for \textsc{RelxillCp} model. The lamp-post geometry gave $\tau=1.30^{+0.30}_{-0.37}$ which is consistent with the \textsc{RelxillCp} model. The observed coronal parameters are consistent with other BH in the LHS \citep{Yan2020}. The height of the corona is obtained as $h\sim 3.4$ $R_g$ and $h\sim 3.7$ $R_g$ from the lamp-post models \textsc{RelxillLp} and \textsc{RelxillLpCp}, respectively. We did not detect any signature of the accretion disk in our analysis of the $0.5-78$~keV Swift/XRT+NuSTAR spectra. We tested this by adding a disk component (\textsc{diskbb} in {\tt XSPEC}), but the f-test rejected this. This indicated that either the disk temperature was very low or the disk normalization was low with a higher disk temperature. However, the source was observed in the LHS, where a low disk temperature is expected \citep[e.g.,][]{RM06}. Hence, a low temperature disk is the most probable reason for non-observation of the thermal emission.

During the spectral analysis, we started the analysis by fixing $q_{\rm in}$ and $q_{\rm out}$ at 3 (\textsc{Relxill-1}). In \textsc{Relxill-2}, \textsc{RelxillCp} and \textsc{RelxillD} models, we allowed $q_{\rm in}$ and $q_{\rm out}$ to vary freely. The outer emissivity index ($q_{\rm out}$) is found to be $\sim 2$ for all three models. We obtained a steeper inner emissivity index ($q_{\rm in}$) with all three models with $q_{\rm in} \sim 6.5-7.2$. A steep inner emissivity profile is expected as the illuminating photons will be largely beamed and focused at the inner accretion flow.

Different flavour and configuration of \textsc{Relxill} model indicated that the accretion disk almost extend up to the ISCO. All models predict the inner edge of the disc, $R_{\rm in} < 1.2$ $R_{\rm g}$. The lamp-post models (\textsc{RelxillLp} and \textsc{RelxillLpCp}) indicated that the disk further moved towards the BH with $R_{\rm in}<1.04$ $R_g$. The inclination angle of the inner accretion flow is obtained to be $i \sim 26^{\circ{}}-37^{\circ{}}$ from the different model. The high density \textsc{RelxillD} model indicated a higher inclination with $i = 37^{+2}_{-3}$ degrees while the lamp-post models indicated a lower inclination with $i\sim 28^{\circ{}}$. The reflection was found to be weak with the reflection fraction ($R_{\rm refl}$) obtained to be $R_{\rm refl} \sim 0.2$ and $\sim 0.16$ from the \textsc{Relxill-2} and \textsc{RelxillD} models, respectively. A different geometry of the lamp-post model yields a stronger reflection with $R_{\rm refl}\sim 0.46$.

The spectral fits with the \textsc{Relxill} and \textsc{RelxillCp} models indicated the iron abundances as $A_{\rm Fe} \sim 2.7-3.2$ $A_{\sun}$. The high density \textsc{RelxillD} model also indicated a similar iron abundances with $A_{\rm Fe} \sim 3$ $A_{\sun}$. The \textsc{RelxillLp} and \textsc{RelxillLpCp} also indicated a similar iron abundances with $A_{\rm Fe} = 3.17^{+0.11}_{-0.13}$ $A_{\sun}$ and $A_{\rm Fe}=3.28^{+0.12}_{-0.11}$ $A_{\sun}$, respectively. It is often observed that the fitting with a low density disk gives a high iron abundances \citep[e.g.,][]{Tomsick2018}. Various studies of Cygnus X--1 with the constant low density model yield the iron abundance $A_{\rm Fe}>9$ $A_{\rm Fe}$ \citep[e.g.,][]{Walton2016,Basak2017}. \citet{Tomsick2018} showed that the same spectra can be fitted with $A_{\rm Fe}=1$ $A_{\sun}$ with the density of the disk as $n\sim 4\times10^{20}$ cm$^{-3}$. In \textsc{Relxill}, \textsc{RelxillCp}, \textsc{RelxillLp} and \textsc{RelxillLpCp} models, the disk density is considered as $n=10^{15}$~cm$^{-3}$ while the density is a free parameter in the \textsc{RelxillD} model. The spectral analysis of \source~ with the \textsc{RelxillD} model returned the disk density as $n < 2\times10^{15}$~cm$^{-3}$. We re-analyzed the data by keeping $A_{\rm Fe}$ fixed at 1. The fit became significantly worse with $\Delta \chi^2>300$ for 1 dof with $n \sim 10^{18}$~cm$^{-3}$. Hence, the high density solution is not required for \source.

The spin of the BH in \source~ is estimated to be very high with $a^* > 0.93$. \textsc{Relxill-1}, \textsc{Relxill-2} and \textsc{RelxillCp} indicated the spin of the BH to be $a^* = 0.97^{+0.02}_{-0.01}$, $0.96^{+0.02}_{-0.02}$ and $0.97^{+0.01}_{-0.02}$, respectively. The \textsc{RelxillD} model even indicated a higher spin, with $a^* > 0.98$. We obtained the spin of the BH as $a^* = 0.97^{+0.02}_{-0.05}$ and $0.95^{+0.01}_{-0.03}$ from the \textsc{RelxilLp} and \textsc{RelxillLpCp} models, respectively. We also checked if a low spin solution is possible for \source. We fixed the spin at 0, and re-analyzed the $0.5-78$~keV XRT+NuSTAR data with all the models. All models returned with a significantly worse fit with $\Delta \chi^2>60$ for 1 dof. Hence, a high spin solution is preferred for \source.

In our analysis, we obtained a good fit from all reflection based models, with $\chi^2_{\rm red}\sim 1.04-1.07$. Statistically, \textsc{Relxill-2} and \textsc{RelxillCp} returned a better fit than other variant of the model with $\Delta \chi^2 \sim 20$. Nonetheless, our analysis shows that GRS~1758--258 hosts a high spinning BH, with $a^*>0.92$. In future, high resolution spetroscopy missions, such as \textit{XRISM} \citep{Tashiro18}, \textit{ATHENA} \citep{Nandra2013}, \textit{AXIS} \citep{Mushotzky2018}, \textit{Colibr\'i} \citep{Heyl2019, Caiazzo2019}, and \textit{HEX-P} \citep{Madsen2018} are expected to constrain or reconfirm the black hole spin of \source~more accurately.

\begin{acknowledgments}
We sincerely thank the anonymous reviewer for constructive suggestions which improved the manuscript significantly. AJ and HK acknowledge the support of the grant from the Ministry of Science and Technology of Taiwan with the grand number MOST 110-2811-M-007-500 and  MOST 111-2811-M-007-002. HK acknowledge the support of the grant from the Ministry of Science and Technology of Taiwan with the grand number MOST 110-2112-M-007-020. AC and SSH are supported by the Canadian Space Agency (CSA) and the Natural Sciences and Engineering Research Council of Canada (NSERC) through the Discovery Grants and the Canada Research Chairs programs. This research has made use of the {\it NuSTAR} Data Analysis Software ({\tt NuSTARDAS}) jointly developed by the ASI Space Science Data Center (ASSDC, Italy) and the California Institute of Technology (Caltech, USA). This work was made use of XRT data supplied by the UK Swift Science Data Centre at the University of Leicester, UK.

\end{acknowledgments}


\vspace{5mm}
\facilities{Swift, NuSTAR.}


\software{\textsc{xspec}; \textsc{nustardas}; \textsc{python}; \textsc{corner.py}; \textsc{astropy}; \textsc{scipy}; \textsc{matplotlib}; \textsc{grace}.}



%\appendix


\bibliography{bhxrb}{}
\bibliographystyle{aasjournal}



\end{document}